# The Crutch or the Ceiling? How Different Generations of LLMs Shape EFL Student Writings

HENGKY SUSANTO, Education University of Hong Kong, China

DAVID JAMES WOO, Everwrite Limited, China

CHINGYI YEUNG, Education University of Hong Kong, China

STEPHANIE WING YAN LO-PHILIP, International Christian School, China

CHI HO YEUNG, Education University of Hong Kong, China

**Abstract.**

The rapid evolution of Large Language Models (LLMs) has made them powerful tools for enhancing student writing. This study explores the extent and limitations of LLMs in assisting secondary-level English as a Foreign Language (EFL) students with their writing tasks. While existing studies focus on output quality, our research examines the developmental shift in LLMs and their impact on EFL students, assessing whether smarter models act as true scaffolds or mere compensatory crutches. To achieve this, we analyse student compositions assisted by LLMs before and after ChatGPT's release, using both expert qualitative scoring and quantitative metrics (readability tests, Pearson's correlation coefficient, MTLD, and others). Our results indicate that advanced LLMs boost assessment scores and lexical diversity for lower-proficiency learners, potentially masking their true ability. Crucially, increased LLM assistance correlated negatively with human expert ratings, suggesting surface fluency without deep coherence. To transform AI-assisted practice into genuine learning, pedagogy must shift from focusing on output quality to verifying the learning process. Educators should align AI functions, specifically differentiating ideational scaffolding from textual production, within the learner's Zone of Proximal Development.

## 1 INTRODUCTION

Large Language Models (LLMs) like GPT-4 [54] and LLaMa [71] are becoming increasingly prevalent, including their utilization to assist English as a Foreign Language (EFL) students in completing writing tasks [59, 67, 68, 80]. Several studies have shown how LLMs can be beneficial for EFL students who face significant challenges in completing their writing assignments, such as limited linguistic skills, a tendency to be less cohesive and fluent, and a higher likelihood of making mistakes [79]. For these reasons, more EFL students are relying on LLMs to complete their writing assignments, especially as these models become more advanced [52].

Some visible practical benefits for EFL students using LLMs include improvements at the sentence level, such as grammar, spelling, proper punctuation, and the use of more sophisticated vocabulary [64, 67, 68, 80]. This is particularly beneficial for less competent EFL writers. Additionally, such advantages were observed in [80] with earlier versions of LLMs prior to the release of ChatGPT in 2023, including the GPT-2 and GPT-1 families. However, the same study also pointed out that LLMs are only helpful to a certain extent; less competent EFL writers cannot completely mask their true writing abilities using earlier versions of LLMs.

As the capabilities of LLMs evolve, number of studies have reported that LLMs released after ChatGPT in 2023 that EFL students benefit from machine-in-the-loop writing, helping them improve their writing composition beyond just sentence-level enhancements. This includes areas such as idea generation, coherence, writing outlines, and organizing thoughts [24, 64]. This raises an important considerations regarding the qualitative differences that more sophisticated LLMs may offer.

Authors' addresses: Hengky Susanto, Education University of Hong Kong, Hong Kong, China; David James Woo, Everwrite Limited, Hong Kong, China; Chingyi Yeung, Education University of Hong Kong, Hong Kong, China; Stephanie Wing Yan Lo-Philip, International Christian School, Hong Kong, China; Chi Ho Yeung, Education University of Hong Kong, Hong Kong, China.

Thus, as LLMs continue to evolve, a crucial question emerges: do these technological advancements enhance their effectiveness—or genuinely render them "helpful" in supporting EFL students' writing? If so, in what specific ways do they provide greater assistance?

In this study, we investigate the extent and limits of LLMs' usefulness as they become more sophisticated, focusing on their role in helping secondary-level EFL students complete writing tasks. This research aims to clarify the degree to which EFL students rely on LLMs and to offer insights into how educators can guide students toward genuine language mastery in the era of AI.

This study investigates the following research questions (RQ):

- RQ-1: What are the benefits for EFL of more advanced LLMs in student writing?
- RQ-2: Which EFL students benefit from more advanced LLMs in student writing?
- RQ-3: How should AI-assisted practice shape the pedagogy strategy in student writing?

To address these research questions, we compare the writing compositions produced by students who received assistance from LLMs both before and after the release of ChatGPT. To achieve this, we collected writing compositions from ninety secondary EFL students in Hong Kong, with roughly half utilizing earlier versions of LLMs released prior to ChatGPT and the other half using more advanced LLMs. The comparison employs two methods: *qualitative* evaluation performed by human assessors and *quantitative* statistical approaches utilizing various readability test methods commonly used in natural language processing (NLP) field [22, 32, 51]. Moreover, this study also a statically analysis of the lexical diversity and sentiment of texts of student compositions.

In the remainder of this paper, we first provide a detailed description of how the samples of student compositions were collected. This is followed by a discussion of the qualitative and quantitative data analyses conducted in this study, which includes an overview of the various tools utilized and developed for the analysis. Next, we present our results followed by a discussion of our findings. The paper will conclude with a discussion of how our findings can be utilized to design strategies to assist EFL students in improving their compositions.

## 2 RELATED WORKS

### 2.1 LLMs in Student Composition

The rapid integration of advanced LLMs like ChatGPT into educational settings has sparked significant research on their impact on student writing, particularly among English Foreign learners [19]. The literature presents a complex picture, highlighting both substantial benefits and potential drawbacks across textual, cognitive, and affective domains. Understanding this nuanced landscape is crucial for effective pedagogical integration and motivates the present study.

A primary focus of this research is how LLMs affect measurable writing quality. Quasi-experimental studies report significant gains in overall scores and in content, organization, vocabulary, and grammar when students use advanced LLMs like ChatGPT [26, 27, 45, 61, 64]. Much of the improvement stems from LLMs' ability to correct surface-level errors (grammar, spelling, basic syntax), acting effectively as proofreaders [2, 3, 61, 67, 80].

Beyond superficial error correction, advanced LLMs seem to push student writing towards greater lexical sophistication and syntactic complexity. Comparative studies note increases in vocabulary diversity, formality, complexity, and the use of specific syntactic structures like adjective modifiers in AI-assisted or AI-generated text [2, 64, 83]. Studies by Zhang & Crosthwaite in [83] found that ChatGPT-3.5 [54] produced text with markedly more formal and complex vocabulary than A second language (L2) writers discussing the same topic. Furthermore, the sheer volume of text produced, combining both human and AI contributions, also correlates with higher quality scores in co-writing scenarios.

However, these measurable improvements often mask deeper concerns, reflecting a potential gap between surface-level metrics and holistic writing quality or "meaning". While adept at surface features, LLMs struggle to identify or address more complex, higher-order errors related to logic, coherence, argumentation, and nuanced meaning, areas where human feedback remains superior [3, 33].

Furthermore, reliance on LLMs may lead to unintended negative consequences. Large-scale analysis of MOOC essays revealed a trend towards less complex and more uniform writing after widespread LLM adoption [43] , a finding echoed by teachers' perceptions of homogenization [46]. This observed trend towards homogenization aligns with neurocognitive findings suggesting LLM-assisted essays are often perceived as less creative and more formulaic [38]. The quality and appropriateness of AI feedback itself can also vary significantly, sometimes failing to align with individual student needs [18].

The cognitive effects of using LLMs are a double-edged sword. On one hand, structured pedagogical interventions have enhanced higher-order skills like critical thinking, problem-solving, and creativity [17], while also bolstering self-regulated learning (SRL) strategies [45]. On the other hand, research warns of "cognitive debt" [38], where reduced mental effort leads to shallower processing and compromised inquiry—especially for users with low prior knowledge [38, 65], potentially hindering the cognitive development essential for writing.

Multiple studies report significant improvements in student motivation, self-efficacy, engagement, and confidence, along with reduced writing anxiety, when using LLMs, particularly for feedback [45, 50, 64, 69, 70]. Students often perceive "AI", which is LLM, as a non-judgmental companion. However, this can be counterbalanced by a potentially diminished sense of ownership over the final product when LLMs contribution is high and student engagement is low [30, 77].

Beyond direct impacts, LLMs offer benefits for the writing process and for providing feedback [50]. Receiving LLM feedback significantly increases student revision behavior [50], and LLMs can provide high-quality feedback when evaluated against pedagogical criteria [26, 27].

While the more advanced LLMs offer potential benefits such as improved textual features [26, 27, 83], enhanced motivation [64, 69], and more efficient feedback processes [50]. However, concerns remain about cognitive debt [38, 65], homogenization [43, 46], decreased creativity [38], diminished ownership [30], and limitations in addressing complex writing issues [3]. This underscores the "Metrics vs. Meaning" gap [81] and highlights the need for research comparing older and more advanced LLMs using both computational metrics and holistic human judgments to capture a comprehensive picture, particularly in the EFL context. Our study aims to address this gap.

## 2.2 The Impacts of LLM in Learning

While studies often report aggregate effects, the impact of LLMs appears to be mediated by learners' diverse characteristics, raising the question of who benefits most. Individual proficiency and prior knowledge play significant roles in human-AI interaction. The negative impact of LLMs on the depth of inquiry was found to be "particularly true for novices" [65]. Baseline EFL studies also observed that AI's impact varies by students' independent writing competence, with lower-proficiency students potentially struggling to leverage AI effectively [79, 80]. Furthermore, feedback engagement rates can differ (E.g., lower in higher grades) [29] , and generic AI support may be less beneficial for struggling writers or those with specific needs [18].

How students engage with LLMs is critical. Lower self-efficacy correlates with heavier reliance on AI and reduced ownership [30], whereas higher metacognitive awareness facilitates more critical engagement with feedback [70]. The ability to effectively prompt the AI is also crucial, with studies revealing diverse interaction patterns and the importance of prompt engineering awareness for strategically directing the AI and overcoming challenges [26, 27, 78]. Methodologies like intent classification [26, 27] and prompt analysis [61] offer ways to diagnose these interaction patterns.

For EFL learners, their linguistic diversity adds complexity. While generative AI offers multilingual capabilities that can support translanguaging, the models often reinforce monolingual norms [31]. Despite this, EFL students actively engage in student-initiated translanguaging using AI tools to support learning [74], suggesting they might interact with AI writing assistants differently than L1 users. Pedagogical frameworks are emerging that advocate for integrating AI to explicitly support these practices [76]. Understanding these dynamics is crucial for determining who and how AI benefits within diverse EFL classrooms.

These existing studies provide valuable insights into how learner characteristics (proficiency, self-efficacy, metacognition, L1 background) and interaction strategies (prompting, translanguaging) influence or mediate students' engagement with AI writing assistants, there is a gap in research that systematically compares older vs. newer LLMs while simultaneously analyzing differential impacts based on EFL writing proficiency (our school banding variable). Our study aims to fill this gap by examining whether the shift to more advanced LLMs exacerbates or alleviates these differential effects, using both textual metrics and human judgments.

### 2.3 Shaping Pedagogy for AI-Assisted Writing

Given the complex benefits, drawbacks, and differential impacts of LLMs, effective pedagogy must strategically guide their use rather than simply providing access.

One of the main problems for pedagogy is that LLMs can lower students' mental effort, potentially leading to less deep thinking and shallower engagement [38, 65] , which may contribute to writing becoming more uniform. Effective pedagogy must therefore actively promote critical engagement, perhaps through structured interventions [17] or by teaching strategic prompt engineering [70, 78]. Furthermore, researchers are exploring new AI tool designs that embed pedagogical strategies directly into the interface, aiming to scaffold revision and reflection while maintaining student authorship, thus encouraging deeper cognitive engagement [35, 36, 62].

Feedback is a key area demanding pedagogical adaptation. As noted previously (see RQ1 discussion), AI excels at identifying surface-level errors [2, 33] , suggesting a strategy of using it as a "first-pass" proofreader, freeing human feedback for higher-order concerns. However, student engagement with AI feedback requires attention, as studies show that many students fail to revise their work after receiving AI suggestions [29]. Additionally, AI feedback quality varies and may lack sensitivity to individual needs , necessitating human oversight and collaborative frameworks [18]. Pedagogy should focus on teaching students to evaluate AI feedback critically, potentially scaffolding metacognitive awareness [69, 70]. Leveraging the positive affective impacts discussed under RQ1 (e.g., increased motivation and confidence) is also key, positioning AI as a non-judgmental "companion". Evaluating the pedagogical quality of AI feedback itself is also crucial.

Pedagogy must consider how LLMs can support the writing process while fostering student agency and voice, perhaps viewing AI as a "creative partner" [14]. Strategies might involve focusing on human-AI dialogue [26, 27] or using AI for lexical/syntactic expansion [67]. Addressing concerns about authorship and ownership [20, 30, 77] requires explicit instruction on ethical use and critical evaluation. LLM integration can also be designed to support self-regulated learning (SRL) strategies [45].

Effective pedagogy must be sensitive to learner differences . Strategies need to account for varying levels of proficiency, providing scaffolding for novices [65, 79, 80] and support for struggling writers [18]. For EFL learners, adopting a translanguaging perspective [31], acknowledging students' student-initiated practices [74], and fostering critical awareness of AI's potential biases are crucial [76].

Teachers' role in this type of interaction remains central, requiring support and training to ensure they can help students maximize the benefits of AI while mitigating its risks ethically [74]. Practical guides [9] and tools like learning analytics dashboards can assist teachers in implementing effective, informed strategies.

These literature point to pedagogical strategies that emphasize critical thinking, reflective behavior, and collaborative human-AI partnerships. Key elements include fostering cognitive engagement, teaching prompt engineering, using AI feedback strategically with human oversight, promoting agency and SRL, addressing ethical considerations, and adapting to learner diversity. However, a gap remains in understanding how these strategies need adaptation based on the specific capabilities of different LLM generations. Our study, by comparing older vs. newer LLMs across proficiency levels, seeks to inform more nuanced, technology-specific pedagogical recommendations.

## 3 METHODOLOGY

In this section, we detail the data collection process and the methods used for analysis.

### 3.1 Data Collection

In this study, we collected samples of writing compositions from 90 secondary school students in Hong Kong between March 2022 and June 2023.

The first phase of data was collected from 44 secondary school students who collaboratively wrote creative stories with an AI chatbot between April and December 2022, prior to ChatGPT's public release [54]. The tool was a prototype text generator built on earlier-generation LLMs—GPT-Neo-1.3B [8], GPT-2 [60], and GPT-J [75]—accessed via an open-source, Hugging Face. At that time, these models typically generated outputs ranging from a single sentence to several sentences in length.

Subsequently, following the public release of ChatGPT, we collected additional data from 46 secondary school students engaged in a feature article writing task (December 2022 – June 2023). A key distinction of this second phase was that students now had access to significantly more advanced and diverse Large Language Models (LLMs). Instead of being limited to a single prototype model, participants used the Poe platform [1], a web and mobile application that aggregates multiple chatbots. Through Poe, students could freely choose among open-source models, including OpenAI's GPT-4 and ChatGPT (GPT-3.5/GPT-4 variants), Anthropic's Claude 3 and 3.5, Google's Gemini series, Meta's Llama models, and others.

To collect data, we conducted two-hour writing workshops at multiple Hong Kong secondary schools representing a range of academic achievement levels. The workshops took place on school premises and followed a consistent structure: students were first taught an explicit EFL writing pedagogy how to write a short story, then introduced them to LLMs and trained in basic prompt engineering techniques.

Students began drafting their compositions during the workshop, after that students were given up to two weeks after the session to revise and submit their final pieces with continued access to the designated LLMs. Students composed their texts using Google Docs and shared the documents with one of the authors. Only compositions of 500 words or fewer were retained for analysis. All workshops were facilitated by the same author, with two additional authors providing support during selected sessions.

To enable transparent analysis of AI contribution, students were instructed to self-report their use of LLMs directly within the document using color-coding: text written entirely by the student was left in black, whereas passages generated or substantially revised with LLM assistance were highlighted in color (students could choose any highlight color for clarity). This approach allowed precise identification of human-authored versus AI-influenced segments.

### 3.2 Methodologies

To analyze whether more advanced LLMs provide greater benefits to EFL students, we adopted two evaluation approaches. First, we employed a set of established readability metrics to objectively measure text complexity and estimated reading difficulty. These metrics offer consistent, standardized, and reproducible assessments, making them particularly suitable for comparing compositions written by different students at different times. Such readability test approaches have been widely used in the natural language processing (NLP) community to evaluate generated and human-written texts [15, 22, 32, 44, 51, 53].

This study employs a comprehensive set of established readability metrics, including the Automated Readability Index (ARI) [63], Coleman-Liau Index [12], Flesch-Kincaid Grade Level [21, 37], Dale–Chall Readability Score [16], Gunning Fog Index [23], Linsear Write [56], Text Standard (an aggregated grade-level estimate derived from multiple indices), and SMOG [47]. All readability scores were computed using the open-source Python library Textstat [4].

A notable exception in score interpretation is the Flesch-Kincaid Grade Level, which presents measurements as negative scores, whereas most other indices assign positive scores to more readable texts.

Despite their effectiveness in measuring surface-level text complexity and structural features, traditional readability metrics may have a critical limitation: they cannot assess semantic coherence or meaningfulness. These formulas rely primarily on proxy variables such as average word length, sentence length, syllable counts, and familiar word lists, without evaluating whether the words form meaningful ideas when combined.

A well-known illustration of this limitation is Noam Chomsky's classic example of a grammatically correct but semantically nonsensical sentence: "Colorless green ideas sleep furiously" [11]. Although this sentence scores as relatively readable under many formulas, the sample sentence makes no real sense, showing why readability scores alone is not sufficient to judge writing quality. A further limitation of traditional readability metrics is that these methods are not sensitive to meaning or conceptual complexity. For instance, the sentence "I think, therefore I am" uses simple vocabulary and grammar, but understanding this sentence may require deeper thinking [53].

Additionally, because the two phases involve LLMs at markedly different stages of technological advancement, a potential concern is catastrophic forgetting (or catastrophic interference) [39]. This phenomenon occurs when further training or fine-tuning of a model on new data or tasks causes it to lose previously acquired knowledge or capabilities. This phenomenon may potentially impact students who utilize more recent LLMs (e.g., GPT-4 and above, Claude 3.5).

To address the limitations of readability metrics, we adopt a second evaluation method consisting of holistic human grading. Each composition was independently scored by one of the authors and an experienced English teacher according to the official Hong Kong Diploma of Secondary Education (HKDSE) English Language paper marking rubric (see Appendix A). Following this widely used scheme in Hong Kong secondary schools, raters assigned separate marks for Content (C), Language (L), and Organization (O). Total $CLO$ score is expressed as $CLO = C + L + O$.

This human evaluation captures semantic and qualitative dimensions that automated metrics cannot assess, including creativity and imagination, story/article development, appropriateness of tone, and coherence of thought. To ensure scoring reliability, the two raters underwent a standardization session before grading and discussed the marking criteria and sample scripts in detail. In cases of disagreement, the final score for each dimension was determined by averaging the two raters' marks.

In addition, the study performs sentiment analysis and assesses the lexical diversity of the writing compositions using several open-source Python libraries for supplementary analyses. Sentiment analysis was performed using the Natural Language Toolkit (NLTK) tool [6]. Lexical diversity was measured with the Measure of Textual Lexical Diversity (MTLD) Python lexical diversity library [41].

In this study, we also employed Pearson's product-moment correlation coefficient [57] to examine the relationships between key variables — for example, the association between readability scores and the proportion of text generated by LLMs versus written independently by students. All correlation analyses were conducted using the built-in statistical functions available in Microsoft Excel [13].

By combining objective readability metrics with expert human evaluation, this study provides a more comprehensive and well-rounded assessment—one that integrates quantifiable technical measures with authentic pedagogical judgment.

# 4 RESULTS AND DATA ANALYSIS

## 4.1 Data Distribution

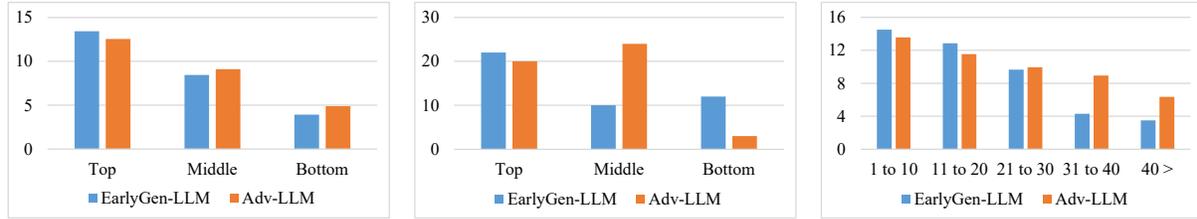

(a) The average of total CLO scores in three categories.

(b) The number of students in three categories.

(c) The average of total CLO scores in groups of five when the total CLO is sorted from highest to the lowest.

Fig. 1. Student performances who received assistant from early generation of LLMs (EarlyGen-LLM) and more advanced LLMs (Adv-LLM) based on the total CLO scores.

To examine whether more advanced LLMs improve overall writing quality as measured by the total CLO score, we first categorized students into three performance groups based on their total CLO scores (out of 15): high or top performers (students who score 10–15 points), middle performers (5–9.99 points), and low performers (below 5 points). We then computed the average CLO score within each band separately for students who used earlier-generation LLMs (Phase 1) and those who had access to more advanced LLMs (Phase 2). An important note is that earlier-generation LLMs generally generate text in sentences or short paragraphs [78]. In contrast, more advanced LLMs can generate the entire articles [79].

As shown in Figure 1a, high-performing students achieved slightly higher average CLO scores with earlier-generation LLMs than with advanced ones. In contrast, mid and low performing students obtained markedly higher CLO scores when using more advanced LLMs. The distribution of students across these groups (Figure 1b) further reinforces this pattern: the proportion of high performers was nearly identical in both phases (approximately 20%–22%), but significantly more students using advanced LLMs were placed in the mid-performing band, and substantially fewer in the low performing group, compared to the earlier-generation LLM group. These findings indicate that more advanced LLMs provide the greatest benefit to mid- and low-performing EFL writers, enabling them to close the performance gap with stronger peers and thereby helping to level the playing field.

One observation worth noting is that high-performing students using earlier-generation LLMs tend to rely less on assistance from these models [80]. However, the situation is mixed for high-performing students using more advanced LLMs: while some of these students still rely less on LLMs, others depend heavily on them for assistance [79]. This observation may suggest that competent writers either require less assistance from LLMs or know how to leverage these models effectively to improve their writing.

To further explore this phenomenon, we ranked all students by their total CLO score and divided them into consecutive groups of 10 (with the exception of the final group, which contained the remaining students). The top-ranked 10 students formed Group 1, the next 10 (ranks 11–20) formed Group 2, and so on. This procedure was applied separately to the earlier-generation LLM cohort (Phase 1) and the advanced-LLM cohort (Phase 2). Within each group of ten (or decile group), we then calculated the mean CLO score. The results are presented in Figure 1c.

This decile analysis confirms the earlier findings: for the highest-performing groups (roughly the top 30%–40%), students using earlier-generation LLMs achieved slightly higher average CLO scores. Starting from approximately the fourth decile onward, however, students with access to more advanced LLMs consistently outperformed their counterparts who used earlier models, with the largest gains observed among the lowest-performing deciles.

Here, we observe that generally the first three groups share similar CLO score. For example, the average CLO score of the top 10 students are similar despite they use earlier or more advance LLMs. Similar pattern is also observed in a group of students who rank between 11 to 20 and 21 to 30 respectively. However, in contrast, we obsere there is a significant gap of the average CLO score where the CLO score of the students who use more advance LLMs is almost two folds of average CLO of the students who use earlier version of LLMS. Similar pattern is also found in the students who rank higher than 40. This further confirms the earlier interpretation that more advance LLMs is more beneficial to lower performer students.

As also shown in Figure 1c, the top three decile groups (ranks 1–30) exhibited highly similar average CLO scores regardless of whether students used earlier-generation or more advanced LLMs. In contrast, starting from the fourth decile (rank 31 onward), a substantial performance gap emerged: students with access to more advanced LLMs consistently achieved markedly higher average CLO scores than those using earlier-generation models. This difference was particularly pronounced in the lower-ranking groups, where the mean CLO score of students using advanced LLMs was often nearly double that of their counterparts using earlier LLMs. These results reinforce our earlier finding that more advanced LLMs provide the greatest benefit to lower-performing EFL students, substantially narrowing the performance gap between low- and mid-performing writers.

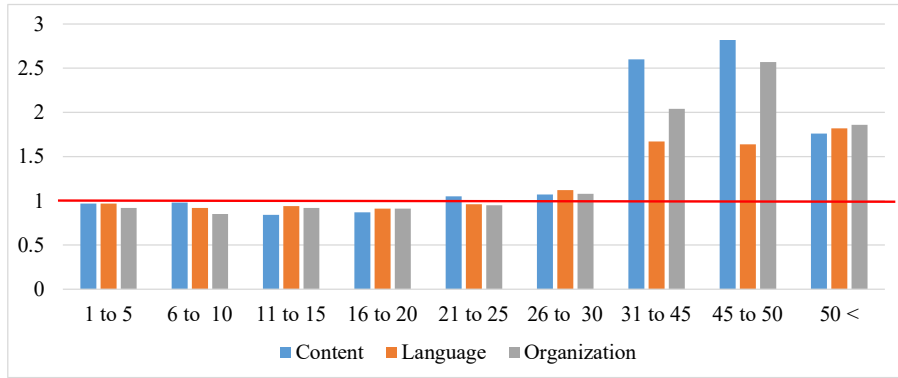

Fig. 2. Improvement of more advance LLMs over the early version of LLMs

## 4.2 What are The Benefits of More Advanced LLMs?

In this subsection, we examine which of the three HKDSE marking dimensions—Content (C), Language (L), or Organization (O) benefits most from the use of more advanced LLMs. To quantify improvement, we define a performance gain score for each dimension as follows

$$C_{improvement} = \frac{C_{advLLMs}}{C_{eLLMs}}, \qquad L_{improvement} = \frac{L_{advLLMs}}{L_{eLLMs}}, \qquad O_{improvement} = \frac{O_{advLLMs}}{O_{eLLMs}}, \qquad (1)$$

where $C_{eLLMs}$, $L_{eLLMs}$, and $O_{eLLMs}$ represent the average Content, Language, and Organization scores, respectively, obtained by students using earlier-generation LLMs, and $C_{advLLMs}$, $L_{advLLMs}$, and $O_{advLLMs}$ denote the corresponding average scores for students using more advanced LLMs. A ratio greater than 1 indicates that advanced LLMs yielded higher average scores in that dimension compared to earlier-generation models. Thus, $C_{\text{improvement}} > 1$ means advanced LLMs were particularly helpful in improving the Content aspect; similarly, $L_{\text{improvement}} > 1$ and $O_{\text{improvement}} > 1$ signal greater benefits for Language and Organization, respectively. This formulation allows us to directly compare the relative impact of advanced LLMs across the three marking dimensions.

To investigate the potential benefit of more advanced LLMS, two sets of students using more advanced and earlier version of LLMs are sorted based their $C, L, O$ scores respectively. Then, compute the performance score with Eq. 1. After that group the improvement score in group of five and we compute the average per group for each of $C, L,$ and $O$ score, as illustrated in 2.

To determine which writing dimension benefits most from more advanced LLMs, we separately ranked students from both cohorts according to their individual C, L, and O scores. For each dimension, we then calculate the improvement ratio for every student using Equation 1. Students are subsequently divided into consecutive groups of five based on their ranked total CLO scores, and the mean improvement ratio is computed for each group in the C, L, and O dimensions separately. The resulting patterns are presented in Figure 2.

This patterns show that the largest gains from advanced LLMs are consistently observed among the lowest-performing students. In particular, these students achieved substantially higher average scores in the Content dimension when using advanced LLMs, followed by notable improvements in Organization and, to a lesser extent, Language, compared to their counterparts who used earlier-generation models. This pattern suggests that more advanced LLMs are especially effective at helping lower-performing EFL writers generate richer content and better-structured compositions, with relatively smaller direct benefits to linguistic accuracy.

## 4.3 Text Complexity Analysis

To complement the human evaluation (The total CLO scores), we also applied a set of established readability metrics to quantify the linguistic complexity of the compositions. These metrics provide a reproducible measure of how easy or difficult a text is to read and understand, based on well-defined mathematical and statistical formulas. Using automated readability scores alongside expert human judgments offers several advantages: *(i)* greater consistency across a large number of compositions, *(ii)* reduced subjectivity, and *(iii)* the ability to reveal patterns (e.g., changes in sentence length, vocabulary difficulty, or syntactic complexity) that may not be feasible in holistic qualitative marking. Thus, the combination of quantitative readability analysis and qualitative expert assessment yields a more comprehensive assessment of writing quality.

In addition to human evaluation, we applied a comprehensive set of established readability metrics previously described (e.g., Gunning Fog Index, Coleman-Liau Index, SMOG Index, Flesch-Kincaid Grade Level, Automated Readability Index, Dale–Chall Readability Score, and Linear Write). These metrics assess multiple surface-level features of the compositions, including the number of sentences, words, and paragraphs; average sentence and paragraph length; the proportion of complex words (typically defined as words with three or more syllables); average syllables per word; percentage of difficult words; and lexical distribution across parts of speech (e.g., percentages of nouns, verbs, adjectives, and their respective difficult forms), among other indicators [10]. This multi-metric approach enabled quantitative characterization of textual complexity and linguistic sophistication in the student writing samples.

Additionally, we examine whether the use of more advanced LLMs leads to differences in the readability profiles of student compositions compared to those produced with earlier-generation LLMs. Then, the study directly compares the scores obtained from each readability metric between the two cohorts.

In comparison analysis, we rank students by total CLO score and divide them into the same three performance bands as before: high or top ( student who scores between 10˜15), mid (5˜9.99), and low (< 5). This stratification allowed direct readability comparisons between the two LLM cohorts at equivalent proficiency levels, similar to the preview section.

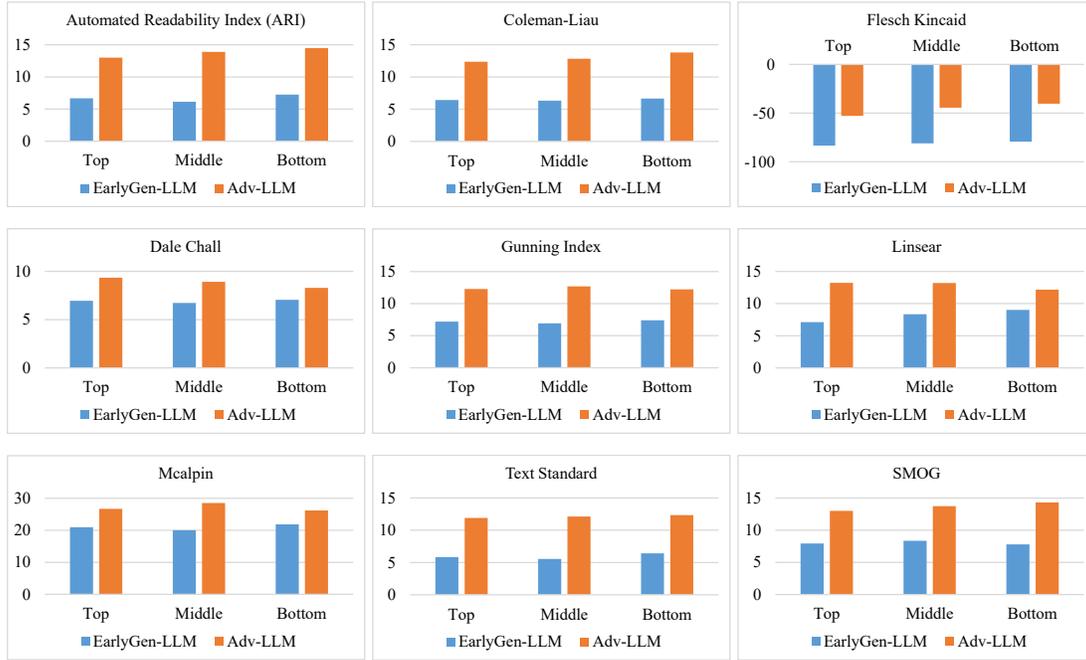

Fig. 3. Readability Test : Automated Readability Index (ARI), Coleman-Liau Index, Flesch-Kincaid Grade Level, Dale-Chall Readability score, Gunning-fog index, Linsear Write, text standard (The estimated school grade level based on Dale-Chall Readability Score) , SMOG

As shown in Figure 3, students who used more advanced LLMs obtained higher readability scores on nearly all metrics than those who used earlier-generation models. However, individual readability metrics differ in their absolute values and relative ranking of the two cohorts because each index emphasizes different textual features. For instance, the Gunning Fog Index heavily weights words with three or more syllables [23], whereas the Coleman-Liau Index relies primarily on average word and sentence length in characters [12]. Despite these methodological differences, the results are highly consistent: across all metrics, compositions produced with the assistance of more advanced LLMs received higher readability scores (indicating more complex) than those written with earlier-generation LLMs.

A further noteworthy finding is that readability scores of more advanced LLMs show no substantial differences across the three CLO performance bands (high, mid, and low). For example, the Coleman-Liau Index scores of high-performing students (top third by CLO) were very similar to those of mid- and low-performing students, and this pattern holds consistently across nearly all readability metrics of more advanced LLM cohorts.

In contrast to the total CLO score analysis, which revealed clear distinctions among high-, mid-, and low-performing students, the readability metrics showed no meaningful differences across these same performance bands. This lack of discrimination suggests that readability test are insensitive to the qualitative dimensions that human evaluators (i.e., English teachers) prioritize and reward in the HKDSE marking scheme, such as richness of content, coherence of ideas, and sophistication of organization.

### 4.4 Analyzing CLO Scores in Relation to LLMs Usage

In the section, we explore the question whether reliant on more advance LLMs will attribute higher total CLO score. Here, CLO is sorted according the percentage or the amount of texts generated by LLMs integrated in the composition. Next, the sorted total CLO are average in a group of five in descending order. For example, the first five CLO scores are averaged together. After that, the next five CLO score will be average together.

The results illustrated in Figure 4 show that there is no obvious pattern indicating whether using more advanced LLMs leads to higher total CLO scores. The figure reveals that some groups of students using earlier versions of LLMs achieve higher total CLO scores, while other groups using more advanced LLMs attain higher scores as well. This can be interpreted as greater reliance on advanced LLMs does not provide a clear advantage, regardless of the student's writing competency. The mixed results suggest that the level of reliance on LLMs may not consistently correlate with the effectiveness of either advanced or less advanced LLMs in helping students completing their writing tasks.

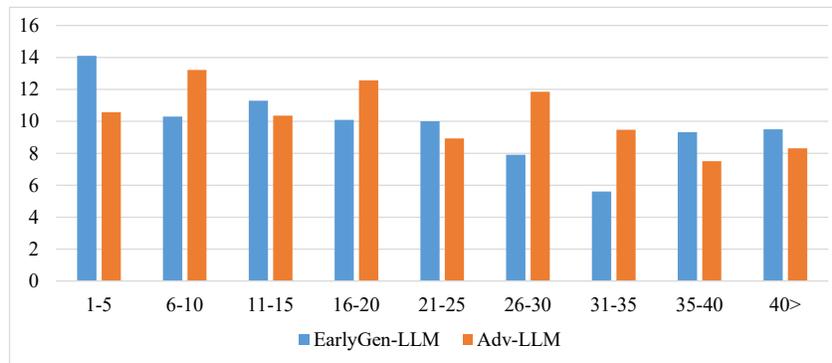

Fig. 4. CLO score sorted according to number of AI text integrated in the writing.

### 4.5 Lexical Analysis

Here, we examine lexical diversity (vocabulary richness) using the Measure of Textual Lexical Diversity (MTLD), a robust sequential metric that is relatively insensitive to text length [48, 49]. Similar to the previous analyses, students were first ranked by their total CLO score and then stratified into the same three performance bands:

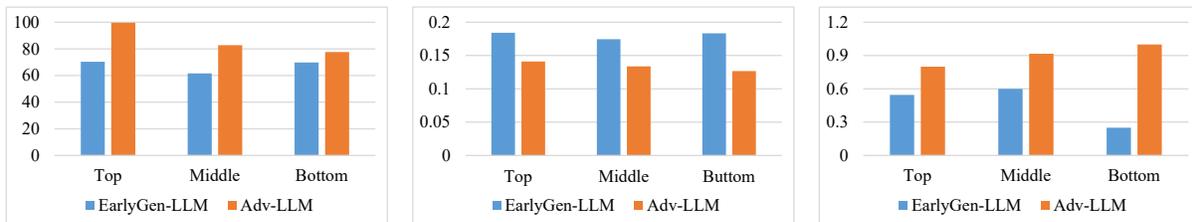

(a) Lexical diversity - MDLT    (b) The percentage of spelling errors    (c) Sentiment Analysis

Fig. 5. Lexical analysis based on the sorted CLO scores.

high (10˜15), mid (5˜9.99), and low (< 5). The average MTLD score was then calculated for each band separately for the earlier-generation LLM and advanced-LLM cohorts.

Our results show that more advanced LLMs significantly improve lexical diversity scores, particularly for students with higher total CLO scores—especially those in the top third. In contrast, the benefits of more advanced LLMs are less pronounced among students with lower total CLO scores, particularly those in the bottom third (Group 3), as shown in Figure 5a.

A notable observation is that the performance gap in lexical diversity between LLM tiers narrows as total CLO scores decrease. This suggests a positive correlation between students' total CLO scores and the extent to which they benefit from more advanced LLMs in terms of lexical diversity. This may suggest that, regardless of the advancement of LLMs, less competent writers may not be capable of leveraging these models to help them write with more diverse lexical variety.

Next, We compare spelling errors in student compositions revised by more advanced LLMs versus earlier models, using the same settings as before. As shown in Figure 5b, advanced LLMs generally reduce spelling errors more effectively. However, this benefit is marginally larger for students with higher CLO scores than for those with lower scores. This can be interpreted as LLMs become more advanced, the benefits of further improvements in correcting spelling errors may follow the law of diminishing returns.

We also analyzed the emotional sentiment expressed in the compositions to determine whether the writer's attitude was positive, negative, or neutral. As shown in Figure 5c, more advanced LLMs generally help students produce compositions with more positive sentiment. The sentiment score gap between advanced and earlier-generation LLM versions increases as students' total CLO scores decrease. This pattern arises because students with lower CLO scores tend to rely more heavily on LLMs to complete their writing tasks, regardless of the model used [79, 80]. Earlier LLM versions typically generate only isolated sentences, whereas more advanced models can produce complete, coherent articles. Consequently, less proficient writers struggle to connect sentences generated by earlier models, but with advanced LLMs they can more easily copy and integrate entire passages into their own work. Since advanced LLMs exhibit a tendency to produce more positively toned text, compositions by these less competent writers also become more positive. This may result in reduced sentiment diversity among the writings of lower-performing students.

## 4.6 Correlation Analysis

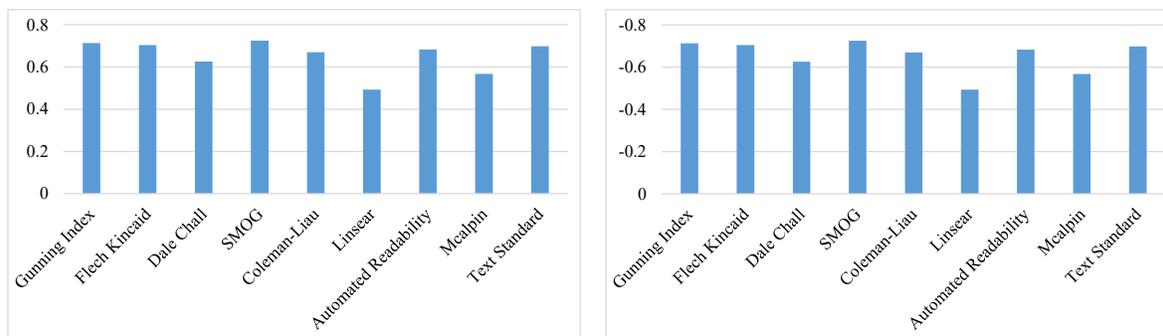

(a) The correlation between number of AI generated words and readability tests.

(b) The correlation between number of Human generated words and readability tests.

Fig. 6. Pearson correlation between readability test and human / AI generated texts

In this subsection, we investigate the strength of the correlation between two evaluation approaches: automated readability tests and human evaluations conducted by the English teachers. These evaluations were applied to both LLM-generated and human-written compositions. To quantify the correlation between the two methods, we employ the Pearson correlation coefficient [57], which is defined as follows.

$$r = \frac{\sum_{i=1}^{n}(x_i - \bar{x})(y_i - \bar{y})}{\sqrt{\sum_{i=1}^{n}(x_i - \bar{x})^2}\sqrt{\sum_{i=1}^{n}(y_i - \bar{y})^2}} \tag{2}$$

where $n$ is the number of observations (student compositions in our case), $x_i$ and $y_i$ are the individual values of the two variables being compared (for instance, readability test scores and human teacher scores), and $\bar{x}$ and $\bar{y}$ represent the sample means of $x$ and $y$, respectively. The score of the correlation coefficient $r$ is $-1 \leq r \leq 1$. A value of $r > 0$ indicates a positive linear correlation (as one variable increases, the other tends to increase), whereas $r < 0$ indicates a negative linear correlation (as one variable increases, the other tends to decrease). When $r = 0$, no linear relationship exists between the variables.

In the first analysis, we computed the Pearson correlation coefficient between readability test scores and the proportion of LLM-generated text in student compositions. This was done by pairing each readability score with the percentage of words generated by the LLM for the corresponding composition. The coefficient $r$ was then calculated using Equation 2. As shown in Figure 6a, there is a moderate to strong positive correlation between the two variables across different readability metrics, with correlation coefficients ranging from approximately 0.7 and 0.5.

In the second analysis, we computed the Pearson correlation coefficient between readability test scores and the proportion of student-written text in each composition. This was achieved by pairing each readability score with the percentage of human-authored words per composition and applying Equation 2. As shown in Figure 6b, the correlation between these two variables is negative.

These findings suggest that readability test scores correlate more positively with LLM-generated texts than with student-written texts. This pattern likely arises because LLMs are trained on large corpora scraped from the internet, which includes academic articles, scientific journals, and other professionally written texts [72]. Such sources frequently employ advanced and complex vocabulary—typically longer, less common words with multiple syllables [82]. Many classic readability tests, including Flesch–Kincaid [37], Flesch Reading Ease [21], and the Gunning Fog Index [23], treat longer words and higher syllable counts as indicators of greater textual difficulty. For example, the word "facilitate" is rated as more complex than "help" largely because it has more syllables and is less frequent in everyday use [23, 82].

In academic and professional writing, such vocabulary is often deliberately chosen for precision and conciseness when expressing nuanced or technical concepts [5, 28]. As a result, when students incorporate LLM-generated passages containing this kind of linguistically sophisticated language, their compositions receive higher complexity scores on automated metrics—even though the students themselves may not possess comparable writing proficiency. As noted in classic readability literature [21, 23, 37], these metrics primarily assess textual difficulty through surface-level linguistic features, such as average sentence length and the proportion of polysyllabic words (typically words with three or more syllables).

In contrast, EFL students' compositions show a negative correlation with readability scores, as less proficient writers tend to use simpler, shorter words. Since readability formulas favor longer words and sentences, student-written texts often result in lower readability (lower complexity) scores.

Next, we examine the correlation between LLM-generated text, student-written text, and their corresponding CLO scores to assess how more advanced LLMs impact student performance. As illustrated in Figure 7, the proportion of LLM-generated text is negatively correlated with total CLO scores, as well as with each individual score for Communication, Learning, and Organization. In contrast, the proportion of student-written text shows a positive correlation with total CLO scores and each of the Communication, Learning, and Organization scores.

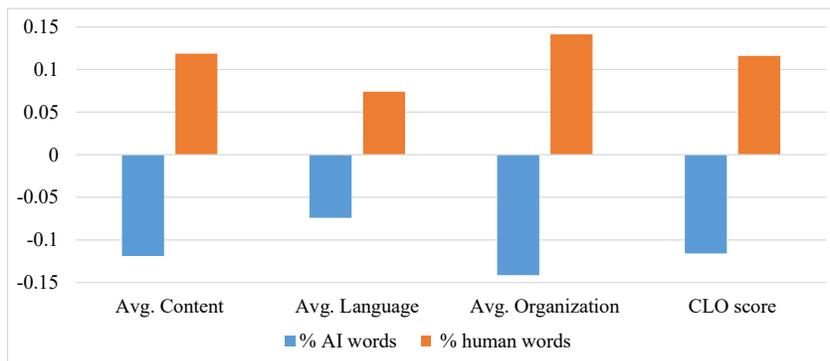

Fig. 7. The correlation between number of AI and human generated words and C, L, and O scores.

As also illustrated in the same figure, the proportion of student-written texts exhibits the strongest positive correlation with the Organization score, followed by the Content score, and then the Language score. This suggests that human evaluators place the greatest emphasis on organization including the logical development of ideas, clear cohesion, and coherent overall structure followed by content quality (development of ideas, relevance, creativity, and imagination). Language mechanics (grammar, spelling, punctuation, and vocabulary choice) are comparatively less influential in determining the total CLO scores (see grading rubric in the Appendix).

These findings further suggest that LLMs are particularly effective at improving language mechanics (grammar, spelling, punctuation, and vocabulary) [64, 80]. However, human evaluators prioritize higher-order skills in EFL writing—specifically, the ability to produce creative, imaginative, well-developed, and contextually relevant ideas that are logically structured and clearly communicated.

## 5 DISCUSSIONS

In this section, we discuss our key findings and propose pedagogical strategies to help EFL students improve their composition writing based on these results.

### 5.1 Findings

Our study shows that as LLMs become more advanced, they increasingly help improve content scores. Earlier findings indicated that LLMs primarily enhance language mechanics [67, 80]. However, more advanced LLMs also boost content quality of composition produce by the lower-proficiency writers, though this benefit plateaus or becomes limited.

Other findings suggest that, in the presence of more advanced LLMs, widely used readability tests may not be suitable for assessing writing quality, particularly in compositions written by EFL students. These metrics reward surface-level complexity (longer words and sentences) typical of LLM-generated text, while failing to evaluate aspects valued by human evaluators, such as creativity, logical organization, and authentic idea development. This limitation is particularly problematic in EFL contexts, where students naturally produce simpler language, yet more advanced LLMs can artificially inflate perceived complexity. As a result, heavy reliance on LLMs may mask less proficient and create a misleading impression of writing ability. Therefore, traditional readability tests may fail to reflect true proficiency in the era of advanced AI assistance.

In our study, human evaluators do not prioritize textual complexity in EFL students who are still developing their language proficiency. Instead, they place greater value on content quality and organization, which better reflect students' critical thinking, creativity, and ability to structure ideas clearly. For language learners, meaningful progress lies in the development of these higher-order skills.

In contrast, LLMs were trained predominantly on professional and published texts (academic articles, literature, legal documents, novels, and domain-specific writing) that feature high linguistic sophistication, longer words, more syllables, and complex sentence structures [25, 54]. As reported in [42, 55, 58], LLMs were trained using "trusted" public data sources like Wikipedia and the reference documents cited in Wikipedia [42]. Other sources include the Books3 platform, which contains over 190000 copyrighted digitized books and was used to train Meta's LLaMA [58]. Google is also reported to have utilized data from its own library of digitized books and the Toronto Books Corpus to train its LLMs [55]. Additionally, OpenAI trained their LLM models on New York Times news articles [58]. These surface-level characteristics are exactly what traditional readability formulas measure and reward, causing LLM-generated text to score highly on such metrics.

This difference stems from the distinct objectives of LLMs and human writers. LLMs are trained to minimize a loss function by accurately predicting the next token, which often results in fluent, statistically probable, and linguistically sophisticated text [25]. While human writers prioritize effective communication, clarity, reasoning, and personal expression.

Moreover, the current LLMs are assessed based on how well they do in standardized tests, such as the US GRE, GMAT, AP, LSAT, SAT, and mathematical exams from the Mathematical Association of America, among others [34]. In contrast, student compositions are holistically judged on criteria such as idea development, higher-order logic, coherence, argumentation, emotional impact, and nuanced meaning [3, 33]. Therefore, the evaluation criteria used for LLMs differ from those used to evaluate compositions written by EFL students. This discrepancy in expectations may highlight the limitations of LLMs in assisting students who may lack the necessary skills to produce high-quality compositions.

Our findings further show that more advanced LLMs enable students—especially lower-proficiency writers who heavily rely on LLM-generated text—to produce compositions with significantly more complex vocabulary. As a result, traditional readability tests strongly favor LLM-influenced writing due to its surface-level sophistication [67], regardless of the students' writing competency. Thus, while advanced LLMs effectively improve mechanical aspects at the sentence level (e.g., grammar, spelling, and punctuation) and can supply content ideas, they do not necessarily elevate writing to a level that is truly unique, evocative, or emotionally resonant.

For these reasons, our study suggests that more advanced LLMs help level the playing field between some low-performing and middle-performing students. These low-performing students may benefit from LLMs' capabilities to generate content and enhance writing mechanics such as grammar, spelling, and punctuation. This may indicate that underperforming students can partially mask their writing abilities with the assistance of LLMs, though only to a limited extent. Thus, even with support from more advanced models, they may not be able to match the writing skills of high-performing students. This is likely because they lack the foundational writing knowledge required to evaluate and refine LLM-generated texts [79, 80].

Finally, as LLMs become more advanced, the sentiment in student compositions tends to become increasingly positive. This shift results in reduced emotional diversity, with most writings adopting a uniformly positive tone. Consequently, heavy reliance on advanced LLMs may lead to compositions that feel less emotionally varied and engaging, which may resulting in monotone and homogeneous writing.

## 5.2 LLMs Impacts to Education

This subsection explore how the utilization of LLMs affects education.

*5.2.1 Navigating Generative AI Affordances.* LLMs hold great potential for learners because they provide readily available affordances [40] that were previously less accessible. Our study highlighted three LLM writing functions that learners used and analyzed how learners at different proficiency levels interact with generative AI:

- *Ideational scaffolding* [65, 80], which refers to information retrieval, synthesis, and argument structuring of texts [65], is primarily supported by the findings in Subsection 4.2, which show that advanced LLMs significantly boost Content and Organization scores for lower-performing writers.
- *Surface-level remediation* [67], which involves linguistic polishing and mechanical correction, is supported by the lexical analysis in Subsection 4.5. The data show that advanced models improve lexical diversity and reduce spelling errors, a function that high-proficiency students used for refinement.
- *Textual production* focuses on the synthesis of original, continuous linguistic sequences in response to a user prompt. This function is supported by the correlation analyses in Subsections 4.4 and 4.6. Those analyses show that lower-proficiency learners often rely on AI to generate coherent text, but excessive dependence can create "cognitive debt" and correlate negatively with human-assigned quality scores, since evaluators prioritize original organization and authorial voice.

This can also include narrative inspiration, plot continuations, and 'creative twists' [80]. Surface-Level Remediation involves polishing the linguistic aesthetics of a draft, such as enhancing lexical sophistication, smoothing syntactic flow, and correcting mechanical errors. Finally, Textual Production focuses on synthesizing original, continuous linguistic sequences based on a user's prompt.

All of these functions can significantly improve the quality and craftsmanship of a text. However, as our study demonstrates, the crux of the issue is whether these functions result in authentic learning gains for the learner. The pedagogical challenge, then, lies in optimizing learning by matching the function to learner proficiency and needs.

Our results suggest that when the function aligned with the learner's ability, authentic learning was more likely. For instance, high-proficiency EFL students leveraged advanced LLMs primarily for lexical refinement and error reduction (Subsection 4.5) whereas lower-proficiency students used AI to mainly support Content and Organization up to a certain level (Subsection 4.2). For advanced writers who have already mastered how to structure text, their needs are in enriching language use and bolstering grammatical accuracy. For beginning writers, their area of support is primarily in generating and sustaining a main topic and organizing the text into one coherent piece. Regardless of their ability level, the students who used AI to match their learning needs, were the ones who most likely experienced the largest learning gains. We argue that this alignment supported learning within their Zone of Proximal Development (ZPD) [73].

*5.2.2 The Human-Computer Interaction Challenge: Mitigating Cognitive Debt .* It is important to mention that for novice learners, the gains in Content and Organization also coincided with higher percentages of AI generated-text (subsection 4.4). Whether this excessive reliance shifts the interaction from scaffolding (guided support) to displacement (cognitive offloading / replacement of learner effort [65] ) is a critical question to investigate in future studies.

Our results shed light on the potential Human-Computer Interaction (HCI) challenges when teachers integrate LLMs into writing instruction. While LLM interfaces are designed to reduce users' cognitive workload, learning requires "'desirable difficulty' - a challenge that improves retention" [7]. Without expert guidance in how to navigate when to use which LLM function, learners are at risk of cognitive debt, where the tool performs the mental labor, leading to shallower processing [38], and ultimately compromising the learner's writing development.

*5.2.3 Implications for Pedagogy.* Although our study did not measure learning outcomes, our results certainly suggest that educators must be proactive in minimalizing the pitfalls of technological misuse. In light of this, we propose three practical strategies for promoting active co-writing when learners use LLMs:

- Match the LLM Function to Learner's ZPD: Instructors should first assess each learner's range of writing skills and level of knowledge. Then, they need to consider how the function will (or will not) scaffold the learner's specific ZPD. Finally, they should carefully select the appropriate set of functions for the learner. For example, if a student struggles with organization, the AI should offer ideational scaffolding rather than textual production [46].
- Teach Prompting: Once the set of functions have been chosen, instructors should explicitly teach students how and when to use them. They should incorporate prompt engineering training so that students learn how to write prompts for the skill or knowledge that they need support with. For instance, if the student struggles with structural planning, then they need to know what prompts are necessary to help them formulate and define clear goals (e.g., argument, tone, structure) [70, 78].
- Instruct AI Output Evaluation: A crucial part in training students how to strategically use AI for both improving the quality of their writing and supporting their learning, is in regularly and consistently providing them guidance and feedback in how to critically evaluate AI suggestions for logic and voice [26, 67]. In addition to direct instruction, previous studies have also suggested that tools like metacognitive checklists for students to assess AI outputs for logic ('Is the reasoning sound?') and voice ('Does it sound personal?') can provide guided practice for students to critically engage with AI [66, 70].

Furthermore, we found a practical limitation to this strategy: Figure 7 reveals a negative correlation between the proportion of AI-generated text and total quality scores. While LLMs excel at mechanics, human evaluators prioritize organization and content, traits strongly correlated with student-written text. Thus, excessive dependency on LLMs not only incurs cognitive debt but also turns learners into passive consumers that produce text lacking the organizational depth and creativity prioritized by human evaluators.

## 5.3 Limitation and Future Works

One of the challenges in this study is scaling up the sample size. A key bottleneck is the involvement of human evaluators. The current study only involves two evaluators, and their grading must be closely coordinated to ensure consistency between the two of them, which is time-consuming. With a larger sample, the evaluation will require more human involvement in assessing the compositions. This necessitates additional coordination among the evaluators to maintain consistency, making the analysis more complex. This issues will be part of our future works.

Each readability metric captures a slightly different aspect of texts, which we did not explore in this study. Future research could investigate the specific insights offered by individual readability tests and their relevance to EFL writing assessment. Understanding how each metric correlates with different writing components—such as coherence, punctuation, and vocabularies usage could provide insights for educators to support EFL students in improving their writing compositions.

One reason why LLMs are very helpful in improving composition at the sentence level is that there are clear rules governing what is correct and incorrect [34]. For example, words must be spelled in specific ways, and grammar must follow established rules. Having these clear rules enables LLMs to generate outputs that meet specific standards. Therefore, EFL students may benefit from LLMs in enhancing their language scores, particularly in areas such as spelling, grammar, and punctuation. However, when it comes to content, while LLMs can be beneficial for EFL students writing non-factual compositions, while they are less suitable for factual writing. One of the limitations of LLMs is their tendency to hallucinate, leading to factual inaccuracies, contextual inconsistencies, and ethical issues [34]. As part of our follow-up work, we plan to investigate how LLMs can be utilized to provide scaffolding for EFL students, guiding them in generating appropriate and accurate content while minimizing the impacts of hallucinations in their compositions.

## 6 CONCLUSION

As the use of AI for collaborative writing among students expands, we investigate how increasingly advanced language models improve student writing and in what specific ways they are beneficial. In this study, we employ both qualitative evaluation by human assessors and quantitative statistical methods. Our results indicate that more advanced LLMs can improve writing quality beyond surface-level mechanics (such as, spelling, grammar, and the use of sophisticated vocabulary), to the improvement of content quality and writing organization. However, this improvement may give the *illusion of competence*, masking the true ability of less skilled writers. Although LLMs improve writing at the level of linguistic mechanics, human evaluators also value organization and content. Thus, to prevent learners from becoming passive consumers who produce texts lacking depth and creativity, we propose three practical strategies based on our findings to encourage active co-writing between EFL students and LLMs. Lastly, as writing is a human-to-human communication medium, our study highlights the role of the human in the loop in guiding students toward authentic expression.

## ACKNOWLEDGMENTS

We very much thank the participating schools, students, and teachers. We would like to Mr. Haoyan Wu for his assistance to improve the graphs.

## 7 APPENDICES

# A  APPENDIX

| Score | Content | Language | Organization |
|---|---|---|---|
| 5 | · Content fulfills the requirements of the question<br>· Almost totally relevant<br>· Most ideas are well developed/supported<br>· Creativity and imagination are shown when appropriate<br>· Shows general awareness of audience | · Wide range of accurate sentence structures with a good grasp of simple and complex sentences<br>· Grammar mainly accurate with occasional common errors that do not affect overall clarity<br>· Vocabulary is wide, with many examples of more sophisticated lexis<br>· Spelling and punctuation are mostly correct<br>· Register, tone and style are appropriate to the genre and text-type | · Text is organized effectively, with logical development of ideas<br>· Cohesion in most parts of the text is clear<br>· Strong cohesive ties throughout the text<br>· Overall structure is coherent, sophisticated and appropriate to the genre and text-type |
| 3 | · Content just satisfies the requirements of the question<br>· Relevant ideas but may show some gaps or redundant information<br>· Some ideas but not well developed<br>· Some evidence of creativity and imagination<br>· Shows occasional awareness of audience | · Simple sentences are generally accurately constructed.<br>· Occasional attempts are made to use more complex sentences. Structures used tend to be repetitive in nature<br>· Grammatical errors sometimes affect meaning<br>· Common vocabulary is generally appropriate<br>· Most common words are spelt correctly, with basic punctuation being accurate<br>· There is some evidence of register, tone and style appropriate to the genre and text-type | · Parts of the text have clearly defined topics<br>· Cohesion in some parts of the text is clear<br>· Some cohesive ties in some parts of the text<br>· Overall structure is mostly coherent and appropriate to the genre and text-type |
| 1 | · Content shows very limited attempts to fulfill the requirements of the question<br>· Intermittently relevant; ideas may be repetitive<br>· Some ideas but few are developed<br>· Ideas may include misconception of the task or some inaccurate information<br>· Very limited awareness of audience | · Some short simple sentences accurately structured<br>· Grammatical errors frequently obscure meaning<br>· Very simple vocabulary of limited range often based on the prompt(s)<br>· A few words are spelt correctly with basic punctuation being occasionally accurate | · Parts of the text reflect some attempts to organize topics<br>· Some use of cohesive devices to link ideas |

Fig. 8.  Assessment Rubric